\documentclass[sigconf]{acmart}
\usepackage{multirow}
\usepackage{enumitem}

\usepackage{array}
\usepackage[table]{xcolor}
\AtBeginDocument{%
  }

\setcopyright{acmlicensed}
\copyrightyear{2018}
\acmYear{2018}
\acmDOI{XXXXXXX.XXXXXXX}
\acmConference[Conference acronym 'XX]{Make sure to enter the correct
  conference title from your rights confirmation email}{June 03--05,
  2018}{Woodstock, NY}
\acmISBN{978-1-4503-XXXX-X/2018/06}

\begin{document}

\title{CityVerse: A Unified Data Platform for Multi-Task Urban Computing with Large Language Models}

\author{Yaqiao Zhu}
\affiliation{%
  \institution{University of Exeter}
  \city{Exeter}
  \country{UK}
}
\email{yz1034@exeter.ac.uk}

\author{Hongkai Wen}
\affiliation{%
  \institution{University of Warwick}
  \city{Coventry}
  \country{UK}
}
\email{hongkai.wen@warwick.ac.uk}

\author{Mark Birkin}
\affiliation{%
  \institution{University of Leeds}
  \city{Leeds}
  \country{UK}
}
\email{M.H.Birkin@leeds.ac.uk}

\author{Man Luo}
\affiliation{%
  \institution{University of Exeter}
  \city{Exeter}
  \country{UK}
}
\email{m.luo@exeter.ac.uk}


\begin{abstract}
  Large Language Models (LLMs) show remarkable potential for urban computing, from spatial reasoning to predictive analytics. However, evaluating LLMs across diverse urban tasks faces two critical challenges: lack of unified platforms for consistent multi-source data access, and fragmented task definitions that hinder fair comparison. To address these challenges, we present CityVerse, the first unified platform integrating multi-source urban data, capability-based task taxonomy, and dynamic simulation for systematic LLM evaluation in urban contexts. CityVerse provides: (1) coordinate-based Data APIs unifying ten categories of urban data—including spatial features, temporal dynamics, demographics, and multi-modal imagery—with over 38 million curated records; (2) Task APIs organizing 43 urban computing tasks into a four-level cognitive hierarchy: Perception, Spatial Understanding, Reasoning \& Prediction, and Decision \& Interaction, enabling standardized evaluation across capability levels; (3) an interactive visualization frontend supporting real-time data retrieval, multi-layer display, and simulation replay for intuitive exploration and validation. We validate the platform's effectiveness through evaluations on mainstream LLMs across representative tasks, demonstrating its capability to support reproducible and systematic assessment. CityVerse provides a reusable foundation for advancing LLMs and multi-task approaches in the urban computing domain.
\end{abstract}



\maketitle

\section{Introduction}
Modern cities generate unprecedented volumes of spatiotemporal data—from real-time traffic flows and crime patterns to satellite imagery and demographic shifts. Harnessing this data to address urban challenges requires solving diverse, interconnected tasks: predicting congestion hotspots, optimizing emergency response routes, recommending service locations, and more. Traditionally, researchers have developed independent, task-specific pipelines, each with custom data processing and modeling choices. This fragmentation limits knowledge sharing, prevents unified reasoning across tasks, and hinders deployment of integrated urban intelligence systems.

The rapid advancement of large language models (LLMs) such as GPT-4\cite{achiam2023gpt} and Gemini\cite{team2024gemini} offers a new opportunity to unify multi-task problem-solving in urban contexts. Unlike traditional models, LLMs excel at cross-domain reasoning, multimodal understanding, and interactive decision-making—capabilities particularly valuable for urban computing where tasks often require integrating spatial knowledge, temporal patterns, and contextual constraints. For instance, answering "Where should we build a new fire station?" demands simultaneous spatial reasoning, demographic analysis, and risk prediction, naturally suited to LLMs' compositional abilities.

Recent efforts have begun exploring LLM applications in urban domains. Benchmarks such as UrBench\cite{zhou2025urbench} and CityBench\cite{feng2025citybench} evaluate models on tasks ranging from geo-localization to spatial reasoning, while systems like CityGPT\cite{feng2025citygpt} investigate text-based urban query understanding. While these studies demonstrate promising potential, they expose a fundamental infrastructure gap: the lack of a unified platform integrating diverse urban data sources, standardized task definitions, and dynamic evaluation capabilities needed for systematic multi-task assessment.

This gap manifests in three critical ways. (i) \textbf{Data accessibility:} Multi-task urban analysis requires simultaneous retrieval of heterogeneous data for a given location—road networks, traffic patterns, demographics, and environmental conditions. Yet existing platforms such as CityFlow\cite{zhang2019cityflow}, SUMO\footnote{https://sumo.dlr.de/docs}, and MATSim\footnote{https://www.matsim.org/} focus on single domains and lack APIs for coordinate-based multi-source access. Researchers must manually integrate datasets from disparate sources, hampering reproducibility and cross-city generalization. (ii) \textbf{Task fragmentation:} Task definitions vary significantly across benchmarks, with inconsistent capability taxonomies. What one benchmark categorizes as "scene reasoning" may overlap with another's "planning" tasks, making fair cross-study comparisons difficult. Without a common framework, it remains unclear whether improvements stem from model advances or task formulation differences. (iii) \textbf{Evaluation limitations:} Current platforms evaluate models on static datasets, lacking simulation environments necessary for testing dynamic decision-making. Real-world urban tasks—such as adaptive traffic signal control or emergency response routing—require closed-loop interaction where decisions affect future states, a capability existing benchmarks cannot assess.


To address these gaps, we present CityVerse, a unified platform integrating multi-source urban data, capability-based task taxonomy, and dynamic simulation for systematic LLM evaluation. CityVerse is the first platform to unify static datasets with interactive simulators under a single spatiotemporal substrate, enabling reproducible assessment across the full spectrum of urban computing tasks—from low-level perception to high-level decision-making.
Our platform makes three key contributions:

\begin{itemize}[leftmargin=6mm]
    \item \textbf{Unified Data APIs:} Coordinate-based access to ten categories of urban data comprising over 38 million curated records, enabling consistent multi-source retrieval across cities without manual integration.
    
    \item \textbf{Capability-Based Task Taxonomy:} A four-level cognitive hierarchy—Perception, Spatial Understanding, Reasoning \& Prediction, and Decision \& Interaction—organizing 43 urban computing tasks with standardized input/output contracts and metrics for fair comparison.
    
    \item \textbf{Integrated Simulation \& Visualization:} Built-in traffic simulation (via CityFlow) coupled with an interactive web frontend, supporting closed-loop evaluation, real-time validation, and intuitive understanding of data distributions.
\end{itemize}


\section{CityVerse Platform}
\subsection{System Overview}
\begin{figure}[t]
  \centering
  \includegraphics[width=0.95\linewidth]{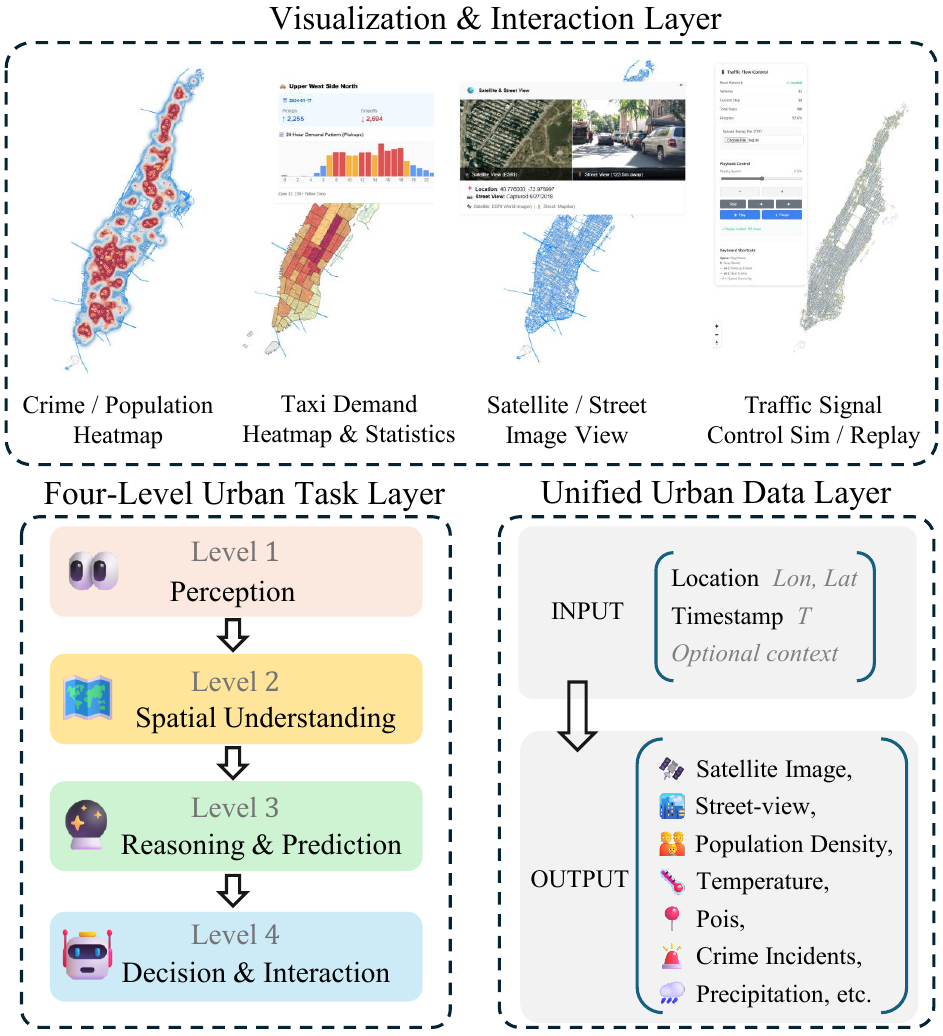}
  \vspace{-1\baselineskip}
  \caption{CityVerse three-layer architecture.}
  \vspace{-1\baselineskip}
  \label{fig:system-overview}
  \Description{The platform unifies heterogeneous urban data, capability-oriented tasks, and interactive visualization through coordinate-based APIs, enabling reproducible multi-task evaluation across cities and models}
\end{figure}

CityVerse is a unified platform designed to enable systematic evaluation of LLMs and other multi-task models on diverse urban computing problems. To address the data fragmentation, task inconsistency, and evaluation limitations identified in Introduction, we adopt a three-layer architecture that decouples data access, task definition, and model interaction through stable, coordinate-based APIs. As illustrated in Figure~\ref{fig:system-overview}, the architecture comprises a data layer providing multi-source urban data, a task layer organizing capability-oriented evaluations, and a visualization layer enabling interactive exploration and validation.

\noindent\textbf{Design Principles.} CityVerse's architecture embodies three core design principles. First, \textit{coordinate-centric unification}: all data and tasks are anchored to a common spatiotemporal substrate, enabling consistent multi-source retrieval without manual integration. Second, \textit{capability-based organization}: tasks are grouped by cognitive complexity rather than domain, establishing standardized input/output contracts and metrics that facilitate fair cross-model comparison. Third, \textit{decoupled extensibility}: data sources, task definitions, and evaluation protocols are separated through stable APIs, allowing new cities, modalities, or tasks to be added without modifying downstream components.

\subsection{Unified Urban Data}




Multi-task urban computing requires simultaneous access to heterogeneous data sources—spatial features, temporal patterns, demographic attributes, and environmental conditions—that traditionally reside in isolated systems with inconsistent schemas and coordinate references. To enable unified retrieval, CityVerse organizes all data around a common spatiotemporal key (location, timestamp) and consolidates the modalities most frequently required across perception, reasoning, and decision-making tasks. Through a systematic survey of prevalent datasets in urban computing literature and practice, we consolidate them into four categories by data modality, as summarized in Table~\ref{tab:urban-data}.

\begin{table}[h]
\begin{center}
\centering
\renewcommand{\arraystretch}{1.2}
\caption{Urban data categories in CityVerse. All temporal data span the full year of 2024 to support longitudinal analysis and seasonal patterns.}
\vspace{-0.5\baselineskip}
\label{tab:urban-data}

\small
\begin{tabular}{>{\centering\arraybackslash}m{1.6cm}>{\centering\arraybackslash}m{2.8cm}>{\centering\arraybackslash}m{3cm}}

\noalign{\hrule height 1.5pt}

\cellcolor{gray!20}\textbf{Category} & 
\cellcolor{gray!20}\textbf{Data Types} & 
\cellcolor{gray!20}\textbf{Primary Sources} \\

\hline

Geographical & POI catalogues; road networks; satellite \& street-view imagery & OpenStreetMap; Google Map; Mapillary API; ESRI World Imagery \\
\cellcolor{black!5}Traffic & \cellcolor{black!5}Traffic flow; vehicle trajectories; signal states & \cellcolor{black!5}City open data portals (e.g. NYC Open Data); simulation outputs \\
Demographic & Population density; Crime records & Census Bureau; city crime portals \\
\cellcolor{black!5}Environmental & \cellcolor{black!5}Temperature, precipitation, wind & \cellcolor{black!5}NOAA stations; local weather services \\

\noalign{\hrule height 1.5pt}

\end{tabular}
\end{center}
\vspace{-1\baselineskip}
\end{table}

\paragraph{Dataset Processing} Data sources are normalized to a unified spatial reference and schema, then processed through an offline-first pipeline to produce both model-ready tables and web-ready tiles. Geographic primitives (roads, buildings, traffic signals) are extracted from OpenStreetMap\footnote{https://www.openstreetmap.org/} using boundary clipping, parsed with geospatial tooling, and exported to GeoParquet/GeoJSON for analysis and MBTiles for rendering. Demographic layers use official census geometries for tract-level population and density; crime records from open city portals are cleaned, geocoded, and time-indexed to support yearly and weekly aggregation. POIs are curated into consistent categories for retrieval and display. Weather data are processed from authoritative stations into daily summaries with standardized fields (e.g., temperature, precipitation, wind). Imagery is handled as a lightweight online extension: satellite tiles are fetched from public basemaps, and street-view thumbnails are retrieved via an API; large assets remain out-of-band to keep the platform portable. Throughout the pipeline, WGS84 coordinates are used for interoperability, UTM projections for metric operations, and uniform spatial units for aggregation. The Data APIs expose a stable contract: given a location and a timestamp, return harmonized features from the relevant categories, enabling reproducible queries across cities and modalities, and providing inputs for both static tasks and dynamic simulation outputs.

New cities or modalities can be added by contributing data sources and registering layers in a manifest without changing downstream interfaces; versioned preprocessing ensures repeatability of benchmarks.


\subsection{Unified Taxonomy of Urban Task}

\begin{figure}[t]
  \centering
  \includegraphics[width=0.85\linewidth]{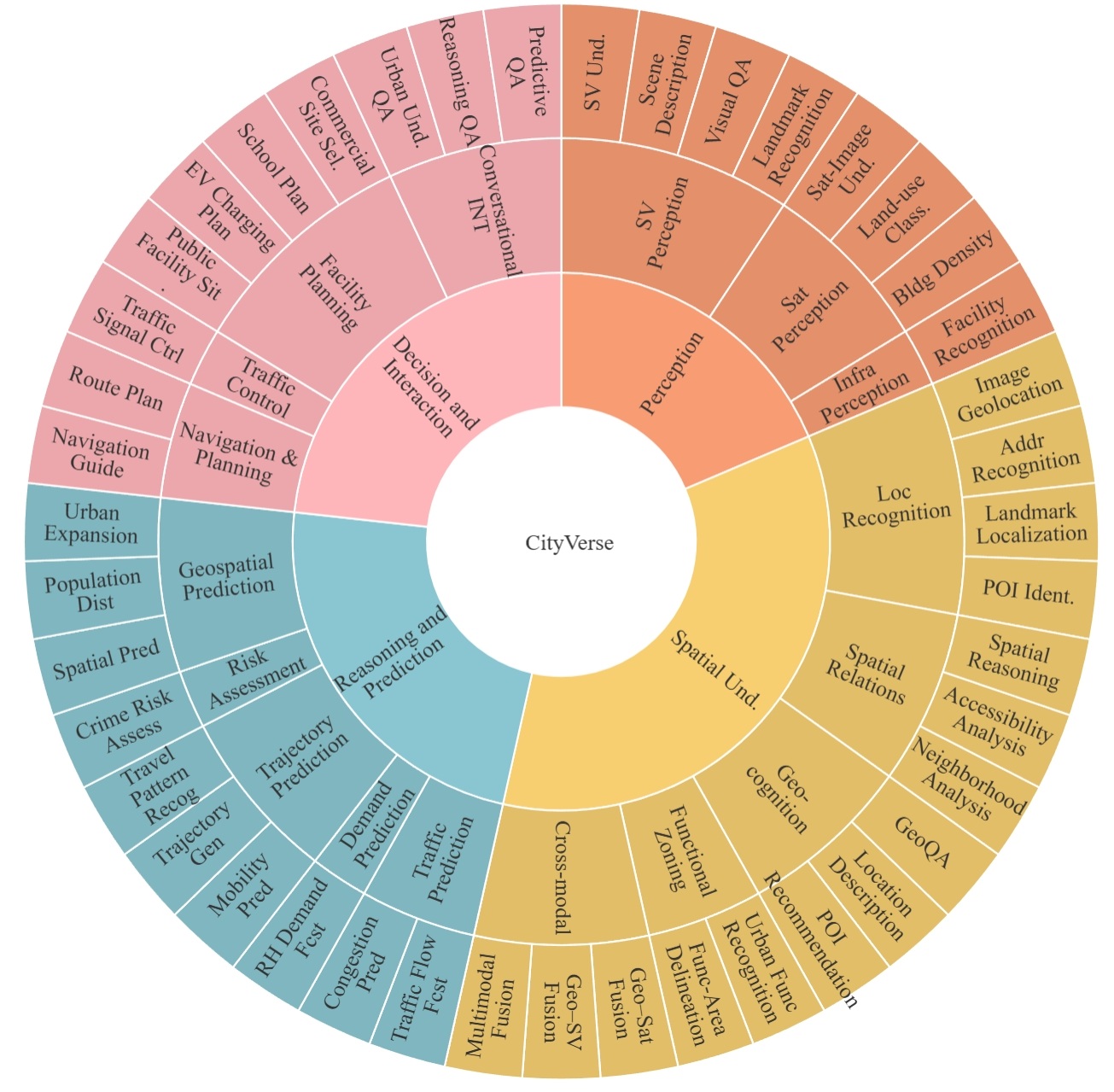}
  \vspace{-1\baselineskip}
  \caption{Unified taxonomy organizing 43 urban computing tasks into a four-level capability hierarchy.}
  \vspace{-1\baselineskip}
  \label{sunburst}
  \Description{
}
\end{figure}

A unified taxonomy is essential for comparable multi-task evaluation with LLMs. Existing urban benchmarks define overlapping or divergent tasks, which obscures capability progress and complicates scoring. Grouping tasks by capability levels provides clear input and output contracts and explicit cognitive assumptions; it aligns prompts, tools, and metrics to the intended difficulty, reduces ambiguity in evaluation, and improves transfer across cities by anchoring tasks to the same spatiotemporal substrate and output schemas exposed by the Data APIs. We therefore organize tasks into four levels that mirror the perception to decision pipeline in urban systems, as shown in Figure~\ref{sunburst}. The levels increase in cognitive complexity while preserving a common interface and metrics, enabling staged difficulty, incremental diagnostics, and fair ablations across models. The four capability levels are as follows:
\begin{itemize}[leftmargin=6mm]
\item \textbf{Perception}: direct extraction from raw inputs (e.g., imagery or text) to identify objects or attributes and produce concise factual outputs without multi-step reasoning.
\item \textbf{Spatial Understanding}: reasoning about proximity, direction, topology, and reachability to generate location-aware judgments grounded in coordinates and map context.
\item \textbf{Reasoning \& Prediction}: using historical or contextual evidence to infer latent causes or forecast near-term outcomes, returning calibrated numeric or categorical estimates.
\item \textbf{Decision \& Interaction}: producing actionable plans or controls, supporting multi-turn refinement and stepwise outputs that enable closed-loop evaluation with a simulator.
\end{itemize}

The task layer abstracts urban problems into model-agnostic Task APIs that bind to data only through spatiotemporal keys. A task invocation takes location, timestamp, and optional context as inputs, queries the Data APIs on demand, and returns structured outputs such as JSON or text. This decouples modeling from storage formats and cities, and it enables reproducible evaluation under consistent inputs. We adopt general metrics that match output types: accuracy and F1 for classification, MAE or RMSE for regression, path quality terms such as length and travel time for planning, and structured matching for key–value or schema-constrained outputs. For decision tasks, the interface supports stepwise interaction with a simulator and standardized replay for fair comparison.

\begin{table*}[t]
\centering
\renewcommand{\arraystretch}{0.8}
\setlength{\tabcolsep}{6pt}
\caption{LLM results on CityVerse tasks.}
\vspace{-1\baselineskip}
\label{tab:llm-results}
\small
\begin{tabular}{lccccccccccc}
\toprule
\multirow{2}{*}{Model} & \multicolumn{3}{c}{Perception} & \multicolumn{4}{c}{Spatial Understanding} & \multicolumn{2}{c}{Reasoning \& Prediction} & \multicolumn{2}{c}{Decision \& Interaction} \\
\cmidrule(lr){2-4} \cmidrule(lr){5-8} \cmidrule(lr){9-10} \cmidrule(lr){11-12}
 & SV ↑ & Sat ↑ & LandUse ↑ & GeoLoc ↓ & GeoQA ↑ & POIRec ↑ & Fusion ↑ & Demand ↑ & Crime ↑ & Nav ↓ & Siting ↑ \\
\midrule
VILA-1.5 3B  & 51.2 & 48.0 & 0.52 & 310 & 0.38 & 0.32 & 0.18 & 0.41 & 0.67 & 5.8 & 12.4 \\
VILA-1.5 13B & 55.0 & 55.7 & 0.58 & 245 & 0.44 & 0.37 & 0.23 & 0.48 & 0.71 & 4.9 & 14.1 \\
VILA-1.5 40B & \underline{62.1} & \textbf{61.8} & \textbf{0.65} & \underline{210} & \underline{0.49} & \underline{0.41} & \underline{0.27} & \textbf{0.58} & \textbf{0.79} & \underline{4.3} & \underline{16.8} \\
LLaVA-1.6 7B & 54.9 & 52.4 & 0.55 & 270 & \textbf{0.52} & 0.35 & 0.21 & 0.46 & 0.70 & 5.1 & 13.7 \\
InternVL2 4B  & 48.7 & 45.5 & 0.48 & 330 & 0.35 & 0.29 & 0.16 & 0.39 & 0.64 & 6.2 & 11.3 \\
InternVL2 8B  & 56.8 & 53.1 & 0.56 & 260 & 0.42 & 0.36 & 0.22 & 0.47 & 0.72 & 5.0 & 14.6 \\
InternVL2 40B & \textbf{64.3} & \underline{60.3} & \underline{0.62} & \textbf{190} & 0.47 & \textbf{0.45} & \textbf{0.30} & \underline{0.55} & \underline{0.76} & \textbf{4.1} & \textbf{18.2} \\
\bottomrule
\end{tabular}
\vspace{-1\baselineskip}
\end{table*}

\subsection{Visualization \& Interaction}
The Visualization \& Interaction layer provides a human-in-the-loop surface to explore unified data, inspect model outputs, and validate tasks while staying aligned with the spatiotemporal substrate exposed by the Data and Task APIs. The web frontend builds on vector and raster tiles in an offline-first manner, using precomputed MBTiles for roads, buildings, population, POIs, and crime, with optional online imagery and street-view for visual context. A manifest-driven loader configures layers, styles, and visibility so that views remain consistent across cities. 


Interaction is organized around location and time. Users can toggle thematic layers to avoid overplotting, use a time slider to examine historical and 2024 crime, and select dates to view weather summaries. Clicking on the map triggers on-demand queries that return attributes for the clicked location and timestamp, surfaced in contextual popups for quick diagnostics. For imagery, satellite basemaps provide background context and nearby street-view thumbnails can be fetched on demand to support perception and spatial understanding tasks, keeping large assets out of band to preserve portability.

CityVerse integrates simulation replay for decision-centric evaluation via a traffic flow view that renders the road network, dynamic vehicles, and traffic signals. Playback controls enable play, pause, step, seek, and speed adjustment; zoom-level rules and coordinate alignment ensure readable overlays that match the real-world map. This view is designed to juxtapose model decisions and observed outcomes, enabling lightweight what-if inspection and transparent, repeatable validation.


\section{Technical Validation and Applications}

We validate CityVerse as a platform, use the NewYork City(NYC) configuration with unified Data APIs and standardized Task APIs. We evaluate mainstream multimodal LLMs under a common prompting and parsing protocol: VILA‑1.5 (3B/13B/40B)\cite{lin2024vila}, LLaVA‑1.6 (7B)\cite{li2024llava}, and InternVL2 (4B/8B/40B)\cite{chen2024far}. 

We align metrics with output types to ensure comparability across models. For perception tasks such as street-view understanding (SV), satellite understanding (Sat), and land-use classification (LandUse), we use accuracy/F1 for categorical labels and exact match(EM) for structured attributes. For spatial understanding, covering image geolocation (GeoLoc), geographic question answering (GeoQA), POI recommendation (POIRec), and multimodal fusion (Fusion), we use localization error (meters or city blocks), EM for QA, and ranking metric Recall@k for recommendation or cross‑modal retrieval. For reasoning and prediction, including ride-hailing demand forecasting (Demand) and crime risk assessment (Crime), we report $R^2$ for regression targets and AUC‑ROC for risk classification. For decision and interaction, namely navigation guidance (Nav) and facility siting (Siting), we assess path quality (length, ETA, constraint satisfaction), coverage and fairness for siting, and constraint violation rates under rule sets shared across models. All prompts, parsing rules, and evaluation scripts are versioned to support exact reproducibility.

The experimental results confirm the effectiveness of the CityVerse platform in enabling multi-level evaluation and reveal clear capability differences across perception, understanding, prediction, and decision layers. Within the same model series, most metrics improve as parameter size increases, showing a consistent scaling trend. InternVL2-40B performs best on POIRec, Fusion, Nav, and Siting tasks, while VILA-40B achieves higher scores on Demand ($R^2$) and Crime (AUC) prediction. LLaVA-1.6-7B shows strong GeoQA ability despite its smaller scale, highlighting the importance of instruction alignment and data composition. Strong perceptual performance, however, does not necessarily imply strong decision-making ability, as cross-layer correlations are not strictly linear. Yet these advanced LLMs perform poorly on multi-view and numerically intensive tasks such as Multimodal Fusion and Ride-hailing Demand Forecasting, which also demonstrates the advantage of our platform in revealing the potential and limitations of LLMs in urban domains.




\section{Conclusion}

CityVerse serves as an extensible foundation for developing next-generation urban AI agents. By decoupling data access, task definitions, and simulation from specific cities or models, the platform enables rapid prototyping and cross-city generalization. We validate the platform through comprehensive evaluation of seven mainstream LLMs, revealing systematic capability gaps between perception and decision-making that highlight areas for future research. 


\bibliographystyle{ACM-Reference-Format}
\bibliography{references}


\end{document}